\documentclass[pre,twocolumn,floatfix,superscriptaddress]{revtex4}
\usepackage{graphicx,amsfonts,amssymb,amsmath, hyperref}
\usepackage[utf8]{inputenc}
\usepackage{multirow}

\newif\ifhyper
\hypertrue
\ifhyper
\hypersetup{
  citecolor = {green},
  colorlinks = {true}, 
  urlcolor = {blue} 
}
\fi

\newlength{\ldag}
\begin{document}

\title{Getting rid of the ghosts: a toy-model of membrane melting}

\author{O. Coquand} 
\email{olivier.coquand@univ-perp.fr}
\affiliation{Laboratoire de Modélisation Pluridisciplinaire et Simulations, Université de Perpignan Via Domitia, 52 avenue Paul Alduy, F-66860 Perpignan, France}


\begin{abstract}
	The theory of thermal fluctuations in crystalline membranes is put under scrutiny.
	In particular, the two critical regimes of the renormalisation group diagram, which are often left out of the discussion because of their instability in one direction,
	are examined in details.
	After studying the proper Goldstone mode counting around each of them, the properties of the fluctuations dominating the large scale spectrum are analysed.
	This shows that the fixed point \textbf{P2} is a good candidate to describe the melting of a crystalline membrane.
	The properties of the melted membrane are then compared to the known properties of fluid membranes.
	As a byproduct of this analysis, we show that the generation of a fluid membrane by melting a bidimensional crystal allows to formulate its correlation functions
	without being plagued by the ghosts that inevitably show up in the usual Canham-Helfrich action relying on the Monge parametrisation.
\end{abstract}

\maketitle

\section{Introduction}

	In the theory of fluctuating surfaces representing physical systems, objects fall into two categories \cite{Nelson04}: \textit{crystalline membranes}
	that encompass the surfaces
	having elastic properties, these are cytoskeletons in biological systems, as well as all the quasi-bidimiensional materials like graphene, h-BN, germanene
	and black phosphorus for example; the second category, called \textit{fluid membranes},
	groups systems for which the motion of constituents along the surface does not cost any energy.
	These are mostly the phospholipid bilayers that form cell walls in living organisms, or more generally liposomes and micelles as soft matter objects.

	If the physics of crystalline membranes is today relatively well understood --- their large scale physics is driven by two different types of symmetry-protected
	Goldstone modes, the phonons that vibrate in the plane of the membrane and the flexurons that vibrate in the orthogonal directions, as a result of which
	the Mermin-Wagner theorem \cite{Mermin66} does not apply \cite{Coquand19}, and the system is at room temperature in an ordered phase called the \textit{flat phase},
	although the general form of the membrane can be any type of locally flat smooth manifold --- the case of their fluid counterpart turns out to be more involved.
	Indeed, though it was established relatively early that in such systems the Mermin-Wagner theorem applies, as a result of which they appear to be flat only
	for length smaller than a characteristic length, called the De Gennes-Taupin length scale $\xi_{GT}$ \cite{DeGennes82}, the rigorous study of fluctuations and
	correlation functions has remained a challenge for quite some time \cite{Gueguen17}.
	This is due to the following simple fact: while for crystalline membranes, the original action is written as a function of a general vector $\mathbf{r}(x)$
	giving the representation of the fluctuating surface as a two dimensional object embedded in three dimensions, for their fluid counterpart, the general way of proceeding
	is by representing it as a height function $h(x)$, the so-called \textit{Monge parametrisation}, which applies very well to any type of surface presenting
	no overhangs, such configurations having a such negligible statistical weight under normal conditions of pressure and temperature that they can be forgotten
	to a very good degree of approximation \cite{David91}.

	The problem comes from elsewhere: it was soon realised that such parametrisation was redundant, there are additional degrees of freedom related by a gauge
	symmetry, and that have to be cancelled away to avoid double counting.
	Such a procedure, which generally relies on the definition of Faddeev-Popov ghosts has been attempted by several authors in the past, but the published
	results were not in agreement with each other \cite{David91,Nelson04,Cai94,Gueguen17,Tonchev21} (we only cite the main contributions here since a complete
	review has been published in \cite{Gueguen17}, to which the reader is referred for more details on the subject).
	The contribution of Gueguen et al., while not in agreement with previously published results, is probably the most rigorous derivation of the ghost action,
	and has not been questioned since its publication to the best of our knowledge.
	Nevertheless, this shows how and why, despite their apparent simplicity, because of the gauge symmetry of the Monge parametrisation, the fluctuations in the
	fluid membranes turn out to be more challenging to describe than their crystalline counterpart.

	In this paper, we propose a toy-model for the fusion of crystalline membranes.
	Thus, the path we follow to land on the fluid membrane's action does not go through the Monge parametrisation, and the problems linked to the
	gauge symmetry and the associated ghost modes do not show up.
	Beyond putting forward a link between the actions of crystalline and fluid membranes, the detailed study of the properties of intermediate actions
	turns out to be very rich in information about how the orientational order can be either sustained (crystalline scenario) or broken (fluid scenario) in theories
	of fluctuating surfaces.

	The paper is organised as follows: we first recall the general properties of the thermal fluctuations in crystalline membranes that will be of use for the following
	study.
	Then, in a second part, we propose a detailed analysis of the spontaneous symmetry breaking patterns, not only at the well-known flat phase fixed point, but at the critical
	points of the crystalline membrane phase diagram.
	The third section is dedicated to the question of the role of fluctuations in the generation of a non-trivial anomalous dimension, linked to the value of the
	lower critical dimension of the model, and thus the applicability of Mermin and Wagner's theorem.
	In the fourth section, we derive the fluid action from the crystalline one through our fusion mechanism.
	Finally, we conclude.

\section{General theory of crystalline membranes at medium temperatures}

	In this first section, we review the main properties of crystalline membranes around the flat phase, without going into all the details of the computations.
	The main contributions are \cite{Aronovitz89,Guitter89,Coquand20,Metayer22} for the perturbative renormalisation group methods, and
	\cite{LeDoussal92,Kownacki09,Braghin10,Hasselmann11,LeDoussal18} for non-perturbative renormalisation.
	Interestingly, a unified picture emerges among all the different methods --- if we put aside some breaking of geometric symmetries by the dimensional
	regularisation \cite{Coquand20}, but this goes beyond the scope of the present study.

	Let us consider the membrane as a $D$-dimensional manifold embedded in a $d$-dimensional space.
	In the following, Latin indices are used for the $D$-dimensional manifold space, and Greek indices for the $d$-dimensional space.
	The theory of elasticity is expressed in terms of strain tensor \cite{Landau90} $\varepsilon$, that we decompose into its irreducible representation of the
	SO$(D)$ group, namely its spin-0 component $\varepsilon^{(0)}$ and its spin-2 component (or deviatoric component in the language of continuum medium mechanics)
	$\varepsilon^{(2)}$, defined as follows:
	\begin{equation}
		\begin{split}
			&\varepsilon^{(0)}_{ij}=\frac{\text{Tr}(\varepsilon)}{D}\delta_{ij} \\
			&\varepsilon^{(2)}_{ij} = \varepsilon_{ij} - \varepsilon^{(0)}
		\end{split}
	\end{equation}

	Under these conditions, the action of crystalline membranes can be written as:
	\begin{equation}
	\label{eqS}
		\mathcal{S} = \int_x\Bigg\{\frac{\kappa}{2}\big(\partial^2\mathbf{r}(x)\big)^2 + K \big(\varepsilon^{(0)}\big)^2
		+\mu \big(\varepsilon^{(2)}\big)^2 \Bigg\}
	\end{equation}
	where $\int_x$ is a shorthand notation for a $D$-dimensional integral, and $\mathbf{r}(x)$ is the position vector on the surface of the membrane.
	It is a $d$-dimensional vector function of the $D$-dimensional variable $x$.
	$K$ is the bulk modulus of the membrane, that characterizes how it responds to compressions and dilations, and $\mu$ is its shear modulus.

	In the following, we are going to work in the $(K,\mu)$ variables since they correspond to the irreducible representation of the strain tensor, and are
	therefore more fundamental, however, in the literature the set of two Lamé moduli $(\lambda,\mu)$ are widely used.
	They are related to our variables through the relation $K = \lambda + 2/D \mu$.

\begin{figure}[h]
		\begin{center}
			\includegraphics[scale=0.5]{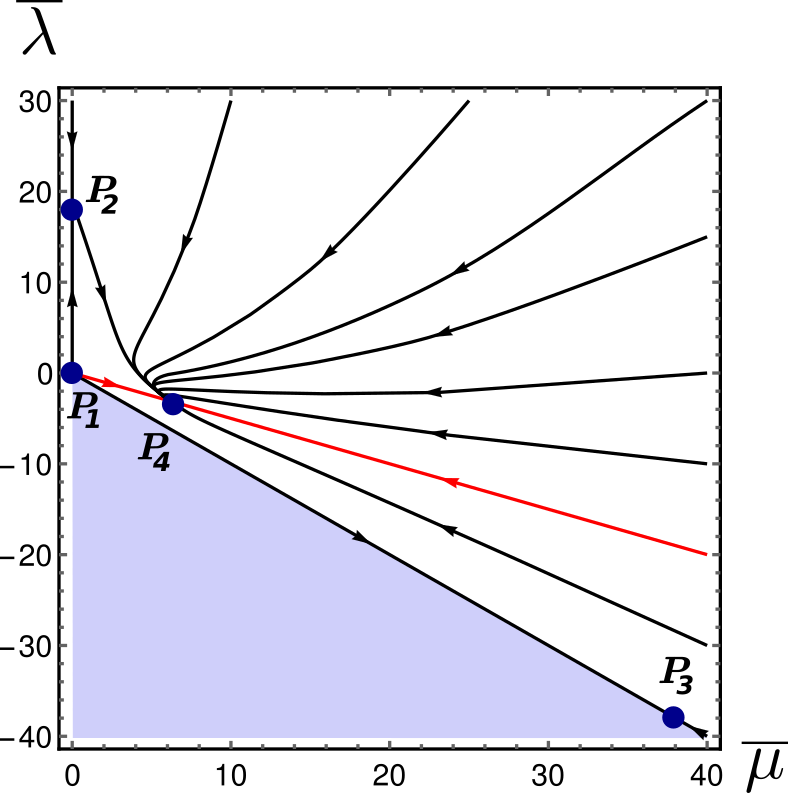}
		\end{center}
		\caption{Renormalisation group flow diagram of crystalline membranes under normal conditions of temperature and pressure.
		The bars above the Lamé parameters indicate that the graph shows only their dimensionless counterpart.
		The shaded area lies outside of the stability region of the elastic potential.}
		\label{figCD}
	\end{figure}

	Under those conditions, the renormalisation group flow of the action (\ref{eqS}) can be computed.
	In the following, we are interested in how fluctuations at small scales modify the measurements of elastic constants that are performed at large scale.
	Hence, we will follow the renormalisation group flow toward the large scale, infrared (IR) limit.
	Independent of the method used to compute it, the general shape of the renormalisation group flow diagram is that of Fig.~\ref{figCD} (for this particular
	picture, the non-perturbative renormalisation group of Mouhanna et al. has been used).
	We shall not discuss here the renormalisation procedures that have been abundantly described in the literature, we refer the reader to the references
	given above for the fundamental papers.
		
	The flow diagram presents four fixed points: (1) the Gaussian fixed point \textbf{P1} $(K=0,\mu=0,\eta_1=0)$ --- $\eta$ being the anomalous dimension ---
	which is unstable in both direction, (2) the flat phase fixed point \textbf{P4} $(K\neq0,\mu\neq0,\eta_4\neq0)$ which is a quasi-universal attractor and describes the
	universal scaling laws of anomalous elasticity of crystalline membranes; it is the presence of such a fixed point that marks the strongest difference between
	crystalline and fluid membranes (for which there is no IR fixed point, so no universality of scaling beyond $\xi_{GT}$), (3) the first critical
	fixed point \textbf{P2} $(K\neq0,\mu=0,\eta_2=0)$ which governs the IR scaling of membranes with a finite resistance to compression/dilation ($K\neq0$)
	but no resistance to shear ($\mu=0$); it is the fixed point that possesses the most common properties with fluid membranes, which is why
	we will call it the \textit{fluid fixed point} in the following, (4) the second critical fixed point \textbf{P3} $(K=0,\mu\neq0,\eta_3\neq0)$, which describes
	membranes with no resistance to compression/dilation, but finite resistance to shear. We will call it the \textit{compressible fixed point} in the following.
	In our opinion, such a material does not really make sense from a physical point of view, but we will nevertheless include \textbf{P3} in our discussion since it
	possesses some properties in common with \textbf{P2} (it is critical with a non-trivial basin of attraction), while sharing others with \textbf{P4} only
	like the existence of a non zero anomalous dimension.
	It will thus serve as an interesting testing platform for our claims.

\section{Identifying the dominant fluctuations: spontaneous symmetry breaking patterns}

	The analysis of Goldstone modes and their counting in condensed matter is more subtle than what the original version of the theorem could let us
	think \cite{Nielsen76}.
	Indeed, it has originally been designed for relativistic systems for which the Lorentz symmetry enforces a linear dispersion relation.
	The publication of an updated counting rule that can be applied to any kind of system is fairly recent in that perspective \cite{Watanabe12,Watanabe14}.
	The updated counting rule also accounts for the fact that the definition of the symmetry breaking mechanism is not sufficient, it is also crucial to examine
	how the broken generators act on the ground state of the system \cite{Low02}.

	With all of this in mind, let us give some definitions: we call $G$ the symmetry group of the action, $H$ that of the ground state (we necessarily have $H\subset G$),
	$Q_a$ a generic broken symmetry generator, and $\rho$ the matrix defined by:
	\begin{equation}
		\rho_{ab}=\left<[Q_a,Q_b]\right>
	\end{equation}
	where $\left<\cdot\right>$ is a shorthand notation for the evaluation in the ground state of the system $\left|0\right>$.
	Under these conditions, the Goldstone modes numbers are given by the following rule \cite{Watanabe14}:
	\begin{equation}
		\begin{split}
			& n_A = \text{dim}(G/H) - \text{rank}(\rho)\\
			& n_B = \frac{\text{rank}(\rho)}{2}
		\end{split}
	\end{equation}
	where to simplify, we will define the type-A Goldstone modes as modes with a linear dispersion relation, and type-B modes as modes with a quadratic dispersion
	relation (the more complicated full definition given in \cite{Watanabe14} boils down to this in our particular case).

	Lastly, let us recall the relations defining the $\mathfrak{iso}(d)$ algebra: if $P_\alpha$ are the generators of translations and $J_{\alpha\beta}$ the
	generators of rotations,
	\begin{equation}
		\begin{split}
			& [J_{\theta\nu},J_{\alpha\beta}] = \delta_{\nu\alpha}J_{\theta\beta} - \delta_{\theta\alpha}J_{\nu\beta} - \delta_{\nu\beta}J_{\theta\alpha}
			+ \delta_{\theta\beta}J_{\nu\alpha}\\
			& [J_{\alpha\beta},P_{\gamma}] = \delta_{\beta\gamma}P_\alpha - \delta_{\alpha\gamma}P_\beta \\
			& [P_\alpha,P_\beta] = 0
		\end{split}
	\end{equation}

	Let us first recall the main results for \textbf{P4}.
	We do not reproduce here the full derivation, the interested reader is referred to \cite{Coquand19}.
	The symmetry breaking mechanism is:
	\begin{equation}
		\text{\underline{\textbf{Mechanism 1:}}}\ \text{ISO}(d)\rightarrow \mathcal{C}\times\text{SO}(d-D)
	\end{equation}
	where $\mathcal{C}$ is the point group of symmetry of the underlying crystalline structure.
	Actually, in case of amorphous graphene for example, or the wrinkled phase of crystalline membranes in presence of quenched disorder
	\cite{LeDoussal18,Coquand18,Coquand20b}, the type of mechanism would still be the same since the Goldstone modes and
	their dispersion relation are determined by the continuous symmetries, which exclude de facto the remaining crystalline symmetries of the ground state
	(the difference in the case of the wrinkled phase appears in the IR, once the anomalous exponents correction to the dispersion relation becomes visible).

	Without going into the details, the symmetry generators of ISO$(d)$ broken by the ground state configuration are:
	\begin{itemize}
		\item The $d-D$ external translations $\big\{P_\alpha\big\}_{\alpha\in[\![D+1;d]\!]}$

		\item The $D$ internal translations $\big\{P_i\big\}_{i\in[\![1;D]\!]}$

		\item The $D\times(d-D)$ mixed rotations $\big\{J_{\alpha i}\big\}_{(\alpha,i)\in[\![D+1.d]\!]\times[\![1;D]\!]}$

		\item The $D(D+1)/2$ internal rotations $\big\{J_{ij}\big\}_{(i,j)\in[\![1;D]\!]^2}$
	\end{itemize}

	Under these conditions,
	\begin{equation}
		\begin{split}
			& \text{dim}(G/H) = 2d - D\\
			& \left<[J_{\alpha_i},J_{\beta_i}]\right>\propto\left<J_{\alpha\beta}\right> = 0\\
			& \left<[P_\alpha,P_\beta]\right> = 0 \\
			& \left<[J_{\alpha i},P_\gamma]\right> \underset{\gamma\neq i,\gamma\neq\alpha}{=} 0 \\
			& \left<[J_{\alpha i},P_\alpha]\right> = \left<P_i\right> \neq 0 \\
			& \left<[J_{\alpha i},P_i]\right> = \left<P_\alpha\right> \neq 0
		\end{split}
	\end{equation}
	so that, it can be established even without knowing the precise value of the coefficients that rank$(\rho) = 2(d-D)$, which leads finally to
	\begin{equation}
		\begin{split}
			& n_A = D \\
			& n_B = d-D
		\end{split}
	\end{equation}
	Hence, crystalline membranes have a spectrum constituted (before renormalisation) of $D$ linear acoustic phonons and $d-D$ quadratic flexurons.

	Let us now consider a membrane at the compressible fixed point \textbf{P3}.
	The ground state has gained a symmetry under global dilations.
	However, because the resistance to shear is still present, the initial pattern of the crystalline lattice is preserved, up to a scale.
	As a result, the number of \textit{independent} symmetry generators acting on the ground state is: $(D-1)$ internal translations (one of them is related to the other
	ones by the dilation symmetry), $(d-D)$ external translations plus $(d-D)-1$ mixed rotations (one is related to the other by the additional internal symmetry).
	As for the matrix $\rho$, it is constructed exactly in the same way, but with the transformation $D\mapsto D-1$.
	Thus,
	\begin{equation}
		\text{dim}(G/H) = 2d - D - 2\ ,\ \text{rank}(\rho) = 2(d - D + 1)
	\end{equation}
	so that
	\begin{equation}
		\begin{split}
			& n_A = D - 1 \\
			& n_B = d - D
		\end{split}
	\end{equation}
	The modes can be identified as follows: $D-1$ is the number of transverse phonons, $d-D$ is the number of flexurons, which means that at the compressible
	fixed point, the longitudinal phonon is not a Goldstone mode anymore.
	This will be indeed confirmed below, where we will show that the longitudinal phonon at the compressible fixed point decouples, and therefore disappears from
	the spectrum.
	Interestingly, this seems to confirm the conjecture of \cite{Coquand19} that the presence of two different types of Goldstone modes is related to the
	existence of a non-trivial anomalous dimension.

	Finally, let us examine the case of membranes at the fluid fixed point \textbf{P2}.
	The absence of resistance to shear allows for internal rearrangements of atoms/molecules in the ground state, which leads to a restoration of the
	ISO$(D)$ invariance in the plane of the membrane:
	\begin{equation}
		\text{\underline{\textbf{Mechanism 2:}}}\ \text{ISO}(d)\rightarrow \text{ISO}(D)\times\text{SO}(d-D)
	\end{equation}
	The broken symmetry generators are:
	\begin{itemize}
		\item The $d-D$ external translations $\big\{P_\alpha\big\}_{\alpha\in[\![D+1;d]\!]}$

		\item The $D\times(d-D)$ mixed rotations $\big\{J_{\alpha i}\big\}_{(\alpha,i)\in[\![D+1.d]\!]\times[\![1;D]\!]}$
	\end{itemize}
	so that,
	\begin{equation}
		\begin{split}
			& \text{dim}(G/H) = d - D\\
			& \left<[J_{\alpha_i},J_{\beta_i}]\right>\propto\left<J_{\alpha\beta}\right> = 0\\
			& \left<[P_\alpha,P_\beta]\right> = 0 \\
			& \left<[J_{\alpha i},P_\gamma]\right> \underset{\gamma\neq\alpha}{=} 0 \\
			& \left<[J_{\alpha i},P_\alpha]\right> = \left<P_i\right> = 0 \\
		\end{split}
	\end{equation}
	Consequently, rank$(\rho)=0$, and
	\begin{equation}
		n_A = d-D\ ,\ n_B = 0
	\end{equation}
	which means that such membranes possess $d-D$ flexurons, but with linear dispersion relation, and no acoustic phonons.
	This is in agreement with what is expected of a fluid membrane.

	All in all, the restoration of the ISO$(D)$ symmetry as a consequence of $\mu=0$ gives to membranes at the fluid fixed point the same IR Goldstone modes as those of fluid
	membranes.
	This is not sufficient to conclude that the two objects are the same.
	In the following, we are going to dive deeper into the properties of crystalline membranes at \textbf{P2} to determine precisely how similar they are to the
	fluid membranes described by the Canham-Helfrich action \cite{Canham70,Helfrich73}.

\section{The lower critical dimension}

	In the O$(N>1)$ model, $D=2$ is the lower critical dimension, which means that the phase transition only takes place at $T=0$.
	It is thus possible to study the model in $D\gtrsim2$ in a temperature expansion to study the properties of the phase transition.
	This procedure involves the definition of a vacuum expectation value for the O$(N)$ field as well as the definition of a non-linear sigma model.

	Such a perspective looks particularly interesting in the case of crystalline membranes where $D=2$ corresponds to the physical dimension.
	Moreover, the following argument, based on mean field theory, gives hopes that $D=2$ could indeed be the lower critical dimension (denoted $D_c^l$ from
	there on).

	In the mean field theory, the dimensions from naive power counting of the elastic moduli are $\Delta(K)=\Delta(\mu)= 4-D$.
	The flexurons propagator behaves in the IR as:
	\begin{equation}
		G_h(q)\underset{q\rightarrow0}{\sim}\frac{1}{q^4}
	\end{equation}
	and that of acoustic phonons is
	\begin{equation}
		G_u(q)\underset{q\rightarrow0}{\sim}\frac{1}{\mu q^2}\sim\frac{1}{q^{6-D}}
	\end{equation}
	(the other acoustic phonon depending on $K$ behaves exactly in the same way).
	Therefore, $D=2$ corresponds to the dimension where the spectrum becomes symmetric, there is no more difference between phonons and flexurons.

	The major flaw of the reasoning above is that it only holds in mean field theory.
	As we have already discussed, at \textbf{P3} and \textbf{P4}, the dimension of the elastic moduli is modified by an anomalous dimension which is particularly strong
	in this model; at either one of these fixed points, the value of $\eta$ is close to 1 \cite{LeDoussal92}.

	Let us dig a bit deeper into this argument.
	We first begin by the theory at \textbf{P4}, as it is the almost universal attractor.
	The following reasoning has first been presented in \cite{David88}.
	In order to develop it, we must restore the $k_BT$ factor that was put equal to one in the action (\ref{eqS}):
	\begin{equation}
		\mathcal{S} = \frac{1}{k_BT}\int_x\Bigg\{\frac{\kappa}{2}\big(\partial^2\mathbf{r}(x)\big)^2 + K \big(\varepsilon^{(0)}\big)^2
		+\mu \big(\varepsilon^{(2)}\big)^2 \Bigg\}
	\end{equation}

	This prefactor can be reabsorbed by the change of variable $\mathbf{r}'=\mathbf{r}/\sqrt{k_BT}$, $\varepsilon'=\varepsilon/k_BT$ (this last one is fixed
	by the relationship between $\varepsilon$ and $\mathbf{r}$ which are not independent fields).
	As a result, the effective couplings become $K_{eff} = K/k_BT$ and $\mu_{eff} = \mu/k_BT$.
	Thus, the effective couplings diverge in the $T\rightarrow0$ limit, which imposes hard constraints on $\varepsilon^{(0)}$ and $\varepsilon^{(2)}$,
	which gives in total 1 + $D(D+1)/2-1$ constraints, or 3 constraints in $D=2$.
	The total number of IR fluctuating modes is $d-D$ flexurons and $D$ acoustic phonons, that is $d=3$ modes.
	Hence the hard constraints in the physical dimensions $d=3$, $D=2$ do not let any mode fluctuate.
	The model is over-constrained and it is not possible to build an expansion around $D=2$.

	This is to be put in regard with the argument of \cite{Aronovitz89}: the long-range order in crystalline membranes is of orientational nature, it is
	therefore typically visible on correlation functions of the type:
	\begin{equation}
		\left<\partial_i\mathbf{h}(x)\cdot\partial_i\mathbf{h}(0)\right>\underset{x\gg a}{\sim} x^{2-D-\eta}
	\end{equation}
	where $\mathbf{h}$ is the flexuron field, and $a$ is the lattice parameter (or any relevant microscopic scale).
	As a result, the lower critical dimension is the solution of the self-consistent equation:
	\begin{equation}
	\label{eqDcl}
		2-D_c^l - \eta(D_c^l) = 0
	\end{equation}

	Since we know that $\eta_4\simeq 1$, $D_c^l<2$, which explains why the expansion around $D=2$ is not possible at \textbf{P4}.

	For the study of the critical fixed points, we are going to need a parametrisation of the strain tensor as a function of the acoustic phonons,
	which are one longitudinal phonon $\mathbf{u}_L$ and $D-1$ transverse phonons $\mathbf{u}_T$.
	The strain tensor is expressed as a function of the metric at the surface of the membrane $g_{ij}=\partial_i\mathbf{r}\cdot\partial_j\mathbf{r}$
	and the metric in the ground state $g^0_{ij}$ as
	\begin{equation}
		\varepsilon_{ij} = g_{ij} - g_{ij}^0
	\end{equation}
	The position vector is decomposed as:
	\begin{equation}
		\mathbf{r} = \mathbf{r}^0 + \mathbf{u}_L + \mathbf{u}_T + \mathbf{h} = \zeta x^i\mathbf{e}_i + \mathbf{u} + \mathbf{h} 
	\end{equation}
	where $\mathbf{r}^0$ is the position vector in the (flat) ground state, and $\{\mathbf{e}_\alpha\}_{\alpha\in[\![1;d]\!]}$ is an orthonormal basis of the embedding space.

	As a result, keeping only the most relevant terms, the strain tensor components become \cite{Aronovitz89,Guitter89}:
	\begin{equation}
		\begin{split}
			& \varepsilon^{(0)}_{ij} = \Big[2\zeta\big(\partial_i\mathbf{u}_L^i\big) + \partial_i\mathbf{h}\cdot\partial_i\mathbf{h}\Big]
			\frac{\delta_{ij}}{D}\\
			& \varepsilon^{(2)}_{ij} = \zeta\partial_i\mathbf{u}_j + \zeta\partial_j\mathbf{u}_i + \partial_i\mathbf{h}\cdot\partial_j\mathbf{h}
			- \varepsilon^{(0)}_{ij}
		\end{split}
	\end{equation}
	Note that while the spin-0 component of $\varepsilon$ only involves the longitudinal phonon (in fact only its divergence), its spin-2 component
	involves only the transverse component; this non trivial fact can be shown by remembering that by definition, the longitudinal phonon has a non
	zero derivative only in one direction, so that the longitudinal terms cancel with the trace when the double contraction of the spin-2 tensor is taken in the
	action.
	The spin decomposition of $\varepsilon$ thus allows for a transverse/longitudinal splitting of the acoustic phonon terms.
	In the perturbative action, when the squares of those terms are developed, the quartic phonon terms, which are sub-dominant, are usually neglected, so that
	only the flexurons possess a four point interaction.
	Also, the kinetic terms from the phonons are defined from the IR dominant terms, so that their coupling to the bending rigidity $\kappa$ also generally
	disappears from the perturbative action.

	We now have all the material to study the fixed point \textbf{P3} in $D=2$.
	At this fixed point, $K=0$.
	We can thus follow the same reasoning as at \textbf{P4}, but now the trace of $\varepsilon$ is no longer constrained.
	The hard constraint imposed as $T\rightarrow0$ are now $(D^2+D-2)/2$, which is $2$ in two dimensions.
	Because $K=0$, there is no coupling to the longitudinal phonon, for which the kinetic term is given by the bending rigidity since no bigger IR term appears in the
	action.
	But, this mode is completely decoupled from the others, and can be integrated over without generating any additional contribution to the action.
	As a result, in physical dimensions, the available fluctuation terms are one flexuron and one transverse phonon.
	Once again, the model is over-constrained, and no perturbative expansion around $D=2$ can be performed.
	Note that this is in agreement with (\ref{eqDcl}) since $\eta_3\neq0$, which suggests a lower critical dimension strictly smaller than two.

	Let us now turn to \textbf{P2}.
	Here, $\mu=0$, so that in the limit $T\rightarrow0$, there is only one hard constraint on the trace of the strain tensor, and this holds independent of the dimension.
	Because $\mu=0$, we have a similar phenomenon as for \textbf{P3}: there is no IR term involving the transverse phonons, which thus appear only in the bending
	rigidity term, decouple from the other modes, and can be integrated over.
	Let us insist that this procedure involves a total integration of the mode, there is no change of variable involved, so in particular no Jacobian and thus no
	Faddeev-Popov ghosts needed.
	There thus remain in physical dimensions two fluctuating modes: a flexuron and the longitudinal phonon, for one constraint.

	Therefore, we can proceed as in the non-linear sigma model: the longitudinal phonon is given a vacuum expectation value, which corresponds to the actual state of dilation
	of the membrane.
	Let us recall that at \textbf{P2}, only the shear modulus is 0, the membrane has a resistance to dilation and compression.
	We thus form the dilaton field $\phi=\partial_i\mathbf{u}_L^i$, since a careful analysis of the spin-0 component of the strain tensor shows that it only
	appears through this combination that is, in fact, a scalar field.
	One can then show that the dilation is a massive field (like the sigma mode in the non-linear sigma model), and the flexuron is the massless Goldstone field,
	analogous to the $\pi$ field.

	All of this is very important since it shows that the model at the fluid fixed point is only stable at $T=0$ in $D=2$, exactly like the model of the
	fluid membrane.
	At $T>0$, the fluctuations keep their well defined scaling properties only up to $\xi_{GT}$.
	In the model at \textbf{P2}, the situation is similar; by analogy with the non-linear sigma model, we can show that the fluctuations keep their scaling
	properties up to a length scale $\xi_2$ which behaves as
	\begin{equation}
		\xi_2 = a \,e^{A/k_BT}
	\end{equation}
	where $A$ is a constant depending a priori on $\kappa$ and $K$.
	The temperature dependence of $\xi_2$ is thus similar as that of $\xi_{GT}$ \cite{DeGennes82}.

	Finally, note that all of this is also consistent with (\ref{eqDcl}) since $\eta_2=0$.

\section{A toy-model action for fluid membranes}

	We can now make a synthesis of all what we have shown above to propose a toy-model for the melting of a crystalline membrane.

	Let us first suppose that, by an unspecified mechanism, the shear modulus of the membrane collapses to 0.
	Here, a true analysis of the fusion conditions, like the proliferation of topological defects, would be required, but this goes beyond the scope of our
	present study.
	Whatever the mechanism, once $\mu=0$, the membrane quits the basin of attraction of the flat phase fixed point \textbf{P4}, and enters that of the fluid
	fixed point \textbf{P2}.

	At this fixed point, there are two main fluctuation modes within the membrane: flexurons, which are the (vector) Goldstone modes,
	and the dilaton, a scalar field, that has become a massive mode, and corresponds to the counterpart of having $K>0$.
	From what we have shown above, the action writes (keeping only the most relevant terms):
	\begin{equation}
		\begin{split}
			\mathcal{S}_{melt}[\mathbf{h},\phi] =& \int_x\Bigg[\frac{\kappa}{2}\big(\partial^2\mathbf{h}\big)^2 + K
				\big(\partial_i\mathbf{h}\cdot\partial_i\mathbf{h}\big)^2 \\
							     &+\frac{Z_\phi}{2}\big(\partial_i\phi\big)^2 + \frac{m_\phi}{2}\phi^2 + g \phi\big(\partial_i\mathbf{h}\big)^2\Bigg]
		\end{split}
	\end{equation}
	with $Z_\phi=\kappa$, $m_\phi=8\zeta^2K$, $g = 4\zeta K$.

	When the membrane is put under well defined conditions of temperature and stress, the dilaton field acquires a vacuum expectation value $\phi_0$.
	In that case, its kinetic term vanishes, and the three point interaction vertex becomes a two point interaction vertex for the flexuron field.
	This calls for two crucial remarks: (i) such a term corresponds precisely to the contribution of the Faddeev-Popov ghosts in the fluid membrane action
	determined by Gueguen et al. \cite{Gueguen17}, (ii) from a more physical perspective, it corresponds to a surface tension term, that can never be put
	to zero in the fluid membrane's action because, even if it is zero at the microscopic scale, it is dynamically generated by renormalisation at intermediate scales.
	This term is also of great importance because it is IR dominant for the Goldstone mode $\mathbf{h}$ and transforms its quadratic dispersion relation into a linear
	one, as predicted by our Goldstone analysis above.

	From the point of view of Mermin-Wagner's theorem, since the dilaton is a massive field, if we were to integrate upon it to get an effective action in terms of
	flexurons alone, the generated effective interaction would be short ranged, as a result of which, the theorem applies, the flat phase at the fluid fixed point
	\textbf{P2} only exists at $T=0$, and at higher temperature, it exists with a reasonable degree of approximation up to the length scale $\xi_2$, in full agreement
	with our earlier findings.

\section{Conclusion}

	All in all, we have shown in this study that the properties of the crystalline membranes at the fluid fixed point \textbf{P2} are similar from all perspectives to
	those of Canham-Helfrich fluid membranes.
	Each of our argument is backed up by an analysis of the other fixed points of the renormalisation flow diagram to ensure the general consistency of what we claim.
	This result is important for two main reasons: (i) by getting to the fluid membrane's model via the crystalline one, we avoid the Monge parametrisation and its
	gauge symmetries; there is no ambiguity in the derivation of the action, and (ii) the Faddeev-Popov ghost contribution arises naturally as the coupling to the dilaton
	field.
	This allows for a physical interpretation of their role in actual materials: the dilaton represents the resistance to dilation/compression stresses and its coupling
	to the flexurons is linked to the bulk modulus of the membrane, a property that was not so easy to read from the Canham-Helfrich action.
	The picture of fluid membranes not just as membranes with no elasticity --- which as we have shown is not totally correct --- should thus be replaced by membranes
	with finite bulk modulus and zero shear modulus.
	From a more formal point of view, this can be reframed as membranes with a spin-0 strain tensor field.

	To upgrade our study from the stage of toy-model to that of actual model of membrane fusion, we would need to precise the actual physical mechanism at play in the
	destruction of the shear modulus.
	This is kept for upcoming studies.

\section{Acknowledgements}

	I thank D. Mouhanna for enlightening discussions.

\bibliography{bibliography.bib}

\end{document}